\documentclass[aps,prd,twocolumn,nofootinbib]{revtex4}

\usepackage{graphicx}
\usepackage{hyperref}
\usepackage{epsfig}
\usepackage{slashed}

\begin{document}

\title{$Z_0$ Boson Decays to $B^{(*)}_c$ Meson and Its Uncertainties}

\author{Li-Cheng Deng, Xing-Gang Wu\footnote{e-mail:wuxg@cqu.edu.cn}, Zhi Yang, Zhen-Yun Fang and Qi-Li Liao}
\address{Department of Physics, Chongqing University, Chongqing 400044, P.R. China}

\begin{abstract}
The programming new $e^{+}e^-$ collider with high luminosity shall provide another useful platform to study the properties of the doubly heavy $B_c$ meson in addition to the hadronic colliders as LHC and TEVATRON. Under the `New Trace Amplitude Approach', we calculate the production of the spin-singlet $B_c$ and the spin-triplet $B^*_c$ mesons through the $Z^0$ boson decays, where uncertainties for the production are also discussed. Our results show $\Gamma_{(^1S_0)}= 81.4^{+102.1}_{-40.5}$ KeV and $\Gamma_{(^3S_1)}=116.4^{+163.9}_{-62.8}$ KeV, where the errors are caused by varying $m_b$ and $m_c$ within their reasonable regions. \\

\noindent {\bf PACS numbers:} 12.38.Bx, 12.39.Jh, 14.40Lb, 14.40.Nd

\end{abstract}

\maketitle

The $B_c$ meson is a double heavy quark-antiquark bound state and carries flavors explicitly. Since its first discovery at TEVATRON \cite{CDF}, $B_c$ physics is attracting more and wide interests. Recently, many progresses have been made for the hadronic production of $B_c$ meson at high energy colliders as LHC and TEVATRON. A computer program BCVEGPY for the direct hadronic production of $B_c$ meson has been presented in Refs.\cite{bcvegpy1,bcvegpy2}. And it has been found that the indirect production of $B_c$ via top quark decays can also provide useful information on $B_c$ meson \cite{tbc1,tbc2,tbc3,tbc4}.

Comparing with the hadronic colliders, an $e^{+} e^{-}$ collider has its own advantages, mainly because of its lower background. As for the previous LEP-I experiment, no $B_c$ events have been found due to its lower collision energy and low luminosity \cite{chang1,chang2}. However, if the luminosity of the $e^+e^-$ collider can be raised up to ${\cal L}\propto 10^{34}cm^{-2}s^{-1}$ or even higher as programmed by the Internal Linear Collider (ILC) \cite{ilc}, then there might have enough events. Moreover, if the $e^{+} e^{-}$ collider further runs at the $Z^0$-boson energy, the resonance effects at the $Z^0$ peak may raise the production rate up to several orders. It has been estimated by Ref.\cite{gigaz} that more than $10^{9\sim 10}$ $Z^0$-events can be produced at ILC per year, which is about $3\sim 4$ orders higher than that collected by LEP-I. Such a high luminosity collider is called as GigaZ \cite{gigaz} or a $Z$-factory \cite{wjw}. Then it will open new opportunities not only for high precision physics in the electro-weak sector, but also for the hadron physics.

The production of $B_c$ through $Z^0$ decays has been studied in Refs \cite{chang1,chang2,braaten} with various methods. Since the process is very complicated, it would be helpful to have a cross check of these results. Furthermore, considering the forthcoming $Z$-factory, it may be interesting to know the theoretical uncertainties in estimating of $B_c$ production.

\begin{figure}
\includegraphics[width=0.40\textwidth]{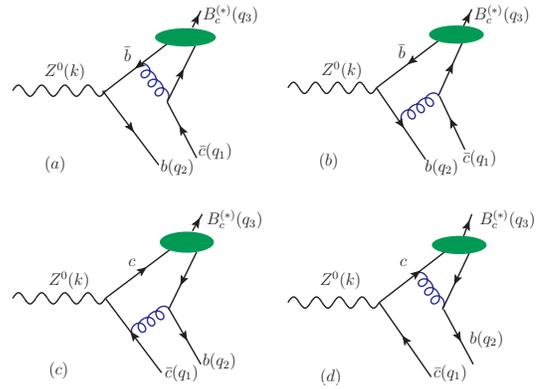}
\caption{Feynman diagrams for the process $Z^0(k)\rightarrow B^{(*)}_c(q_3) + b(q_2) +\bar c(q_1)$. } \label{feyn}
\end{figure}

For the purpose, we need to calculate the process $Z^0 \to B^{(*)}_c+b+\bar{c}$, whose Feynman diagrams are shown in Fig.(\ref{feyn}). According to the NRQCD factorization formula \cite{nrqcd}, the decay width for the process $Z^0 \to B^{(*)}_c+b+\bar{c}$ can be written in the following factorization form:
\begin{equation}
d\Gamma=\sum_{n} d\hat\Gamma(Z^0 \to c\bar{b}[n]+b+\bar{c})\langle{\cal O}^H\rangle,
\end{equation}
where the matrix element $\langle{\cal O}^{H}(n)\rangle$ is proportional to the inclusive transition probability of the perturbative state $c\bar{b}[n]$ into the bound states of $B_c$. As for the two color-singlet $S$-wave states $c\bar{b}[^1S_0]$ and $c\bar{b}[^3S_1]$, their matrix elements $\langle{\cal O}^{H}(n)\rangle$ are related with the Bethe-Salpeter wave function at the origin that can be determined by the potential model \cite{pot1,pot2,pot3,pot4,pot5,pot6}. $d\hat\Gamma(Z^{0}\to c\bar{b}[n]+b+\bar{c})$ stands for the short-distance decay width, i.e.
\begin{equation}
d\Gamma(Z^{0}\to c\bar{b}[n]+b+\bar{c})= \frac{1}{2k^0} \overline{\sum}  |M|^{2} d\Phi_3,
\end{equation}
where $\overline{\sum}$ means we need to average over the spin states of initial particles and to sum over the color and spin of all the final particles. And in $Z^0$ rest frame, the three-particle phase space can be written as
\begin{displaymath}
d{\Phi_3}=(2\pi)^4 \delta^{4}\left(k_0 - \sum_f^3 q_{f}\right)\prod_{f=1}^3 \frac{d^3{q_f}}{(2\pi)^3 2q_f^0}.
\end{displaymath}

The hard scattering amplitude for the process $Z^0(k)\rightarrow B^{(*)}_c(q_3) + b(q_2) +\bar c(q_1)$ can be written as:
\begin{equation}
iM = {\cal{C}}{\bar u_s}({q_2}) \sum\limits_{n = 1}^4 {A _n }{v_{s'}}({q_1}),\label{MM}
\end{equation}
where $\cal{C}$$=\frac{e g_s^2}{\sin\theta_{w}\cos\theta_w}\times \frac{4}{3\sqrt{3}}$. The gamma structure $A_n$ ($n=1$, $\cdots$, $4$) corresponds to the four Feynman diagrams in Fig.(\ref{feyn}). More explicitly, $A_n$ can be written as
\begin{eqnarray}
A_1 &=& \left[ {{\slashed{\epsilon}(k)}{\Gamma_{z\bar b}}\frac{\slashed{q}_2 - \slashed{k} + {m_b}}{(q_2 - k)^2 - m_b^2}{\gamma_\rho} \frac{\chi^{S{S_z}}_{q_3}(q)} {(q_{31} + {q_1})^2}{\gamma_\rho}}\right]_{q=0}, \label{A1}\\
A_2 &=& \left[\gamma_{\rho}\frac{\slashed{k}-\slashed{q}_{32}+{m_b}}{(k-q_{32})^2- m_b^2}{\slashed{\epsilon}(k)}{\Gamma_{z\bar b}} \frac{\chi^{S{S_z}}_{q_3}(q)} {(q_{31} + {q_1})^2}{\gamma_\rho}\right]_{q=0}, \label{A2}\\
A_3 &=& \left[\gamma_{\rho}\frac{\chi^{S{S_z}}_{q_3}(q)}{({q_{32}}+ {q_2})^2} \slashed{\epsilon}(k) \Gamma_{zc} \frac{\slashed{q}_{31} - \slashed{k} + m_c}{(q_{31} - k)^2- m_c^2}{\gamma_\rho}\right]_{q=0} \label{A3}
\end{eqnarray}
and
\begin{eqnarray}
A_4 &=& \left[\gamma_{\rho}\frac{\chi^{S{S_z}}_{q_3}(q)}{(q_{32} + {q_2})^2} \gamma_{\rho}\frac{\slashed{q}_3 + \slashed{q}_2 + {m_c}}{({q_3} +{q_2})^2 -m_c^2} {\slashed{\epsilon}(k)}{\Gamma_{zc}} \right]_{q=0}, \label{A4}
\end{eqnarray}
where $\Gamma_{z\bar b} =\frac{1}{4} - \frac{1}{3}\sin^2\theta_w - \frac{1}{4}\gamma^5$, $\Gamma_{zc}=\frac{1}{4} - \frac{2}{3}\sin^2\theta_w -\frac{1}{4}\gamma^5$ and $q$ is the relative momentum between the two constitute quarks of $c\bar{b}$-quarkonium. In the nonrelativistic approximation, the $S$-wave projector $\chi^{S{S_z}}_{q_3}(q)$ takes the following form
\begin{equation}
\chi^{S{S_z}}_{q_3}(q)= \frac{-\sqrt{m_{B_c}}}{{4{m_b}{m_c}}}(\slashed{q}_{31}- m_b) \left({\alpha\gamma_5+\beta\slashed{\varepsilon}_s}(q_3)\right) (\slashed{q}_{32} + m_c)
\end{equation}
where $\alpha=1(0)$ and $\beta=0(1)$ for $S=0(1)$ meson respectively. $\varepsilon_s({q_3})$ is the polarization vector, $q_{31}$ and $q_{32}$ are the momenta of the two constitute quarks of the meson,
\begin{equation}
q_{31} = \frac{m_b}{m_{B_c}}{q_3} + q \;\;{\rm and}\;\;
q_{32} = \frac{m_c}{m_{B_c}}{q_3} - q,
\end{equation}
where $m_{B_c}= m_b + m_c$ is implicitly adopted to ensure the gauge invariance of the hard scattering amplitude.

By using the conventional trace technique, we need to derive the squared amplitude, which is very complicated and lengthy for the present case. To derive analytical expression for the process $Z^{0}\rightarrow B^{(*)}_c + b +\bar{c}$ and to make its form simpler as much as possible, we adopt the `new trace amplitude approach' suggested and developed by Refs.\cite{chang1,tbc2} to do our calculation. Under the approach, we first arrange each of the four amplitudes listed in Eqs.(\ref{A1},\ref{A2},\ref{A3},\ref{A4}) into four orthogonal sub-amplitudes according to the four spin combinations of the outgoing $b$-quark and $\bar{c}$-antiquark, and then do the trace of the Dirac-$\gamma$ matrix strings at the amplitude level by properly dealing with the massive spinors, which results in explicit series over several independent Lorentz-invariant structures. And then, our task left is to determine the coefficients of these Lorentz-invariant structures. To make the paper more compact, we present the detailed formulae for dealing with the process in Appendices A and B, where Appendix A gives the phase-space integration formulae and Appendix B gives the `new trace amplitude approach', which presents all the necessary coefficients for the Lorentz-invariant structures.

As a cross check of the present obtained results, by taking the same parameters, we can obtain consistent numerical results as that of Ref.\cite{chang1} within reasonable numerical errors \footnote{There are some typos in the formulae listed in the Appendix of Ref.\cite{chang1}, the right ones are presented in the present Appendix B. }.

In doing the numerical calculation, we take $m_Z=91.1876$ GeV and $\alpha_s(m_Z)=0.1176$ \cite{pdg}. To be consistent with the present leading-order calculation, we adopt the leading-order $\alpha_s$ running, and by taking the normalization scale to be $2m_c$, which leads to $\alpha_s(2m_c)=0.212$. The two constitute quark masses are taken as $m_b=4.90$ GeV and $m_c=1.50$ GeV. With the above parameter values, it can be found that the total decay width $\Gamma_{(^1S_0)}=81.4$ KeV and $\Gamma_{(^3S_1)}=116.4$ KeV.

\begin{figure*}
\includegraphics[width=0.40\textwidth]{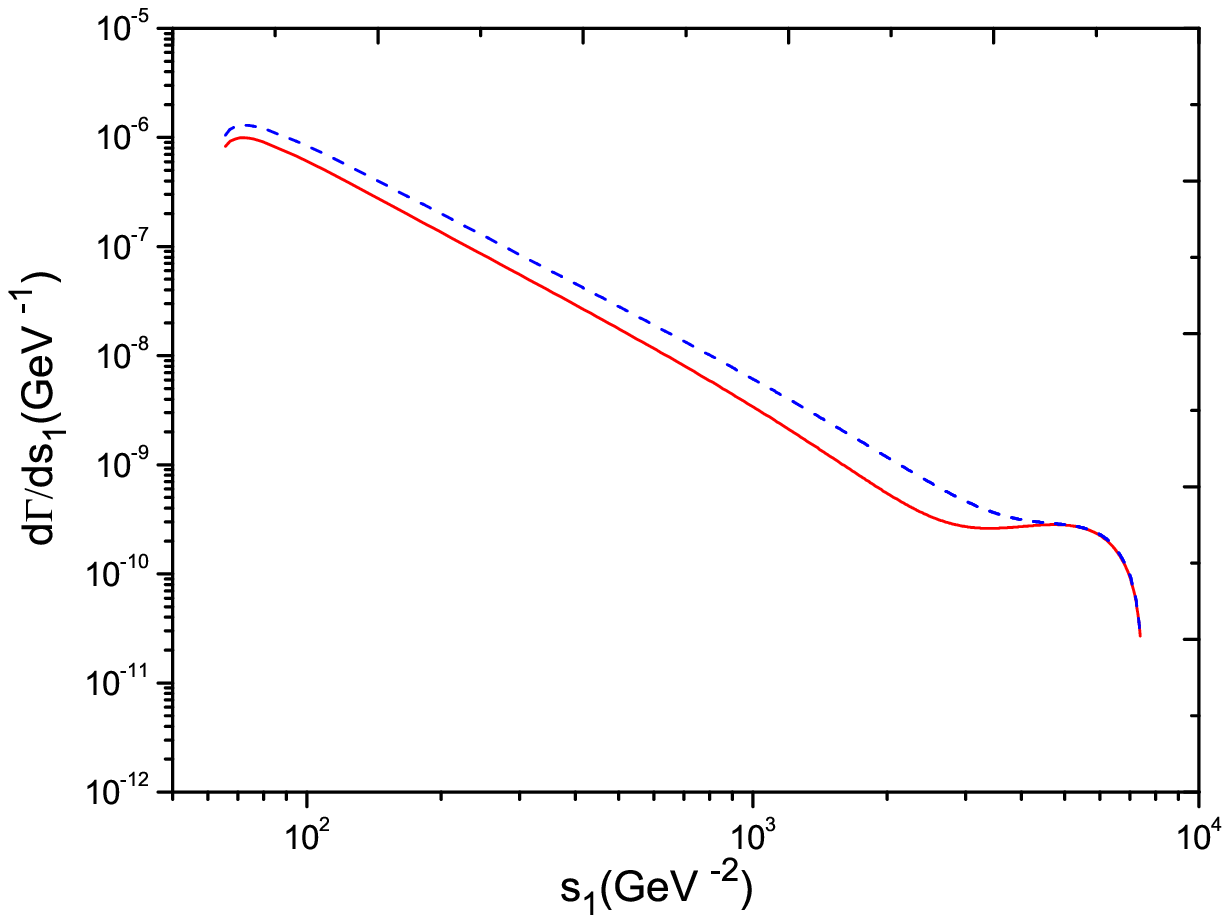}
\includegraphics[width=0.40\textwidth]{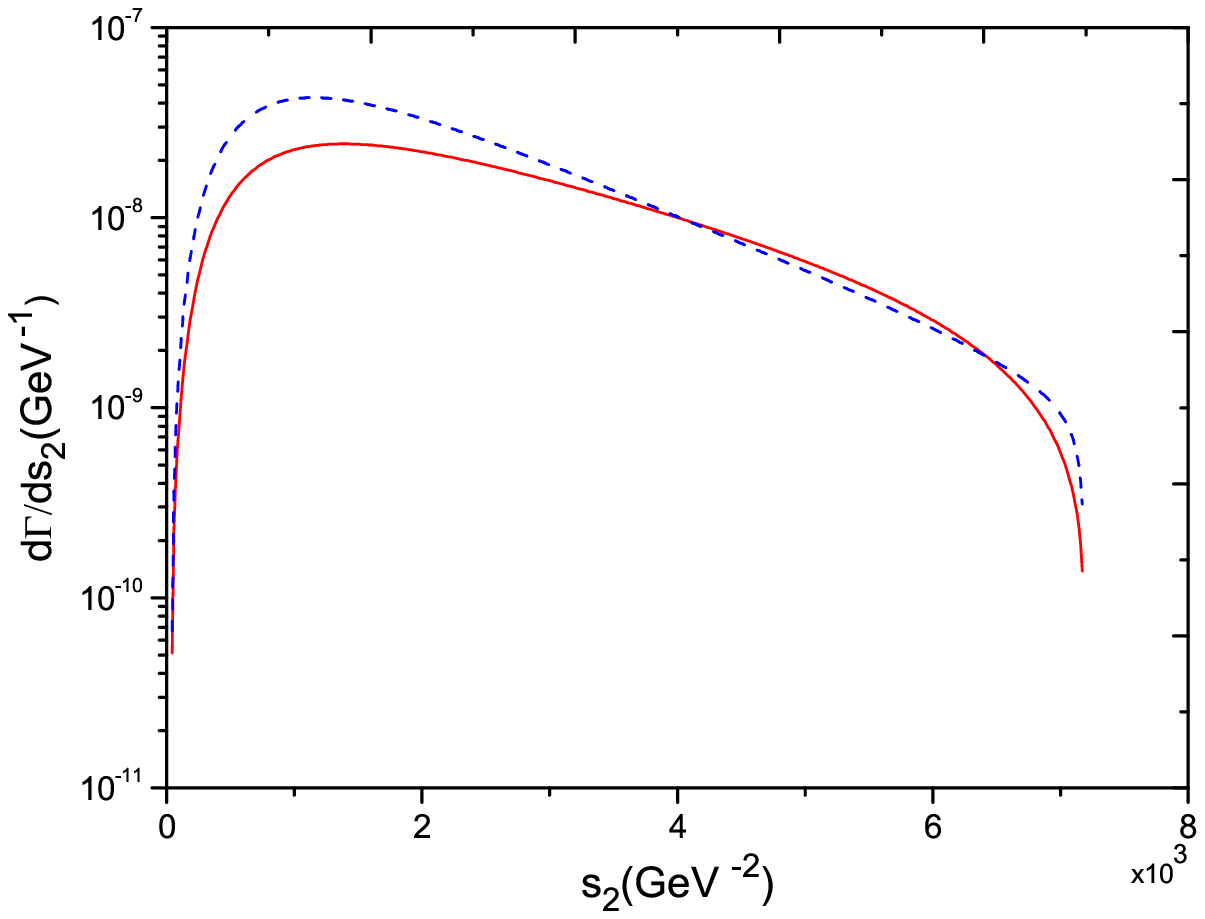}
\caption{Differential decay width $d\Gamma/ds_1$ (Left) and $d\Gamma/ds_2$ (Right) for the process $Z^0\rightarrow B^{(*)}_c+b+\bar{c}$, where the solid and the dashed lines are for $B_c$ and $B^*_c$ states, respectively.} \label{diss1s2}
\end{figure*}

\begin{figure*}
\includegraphics[width=0.40\textwidth]{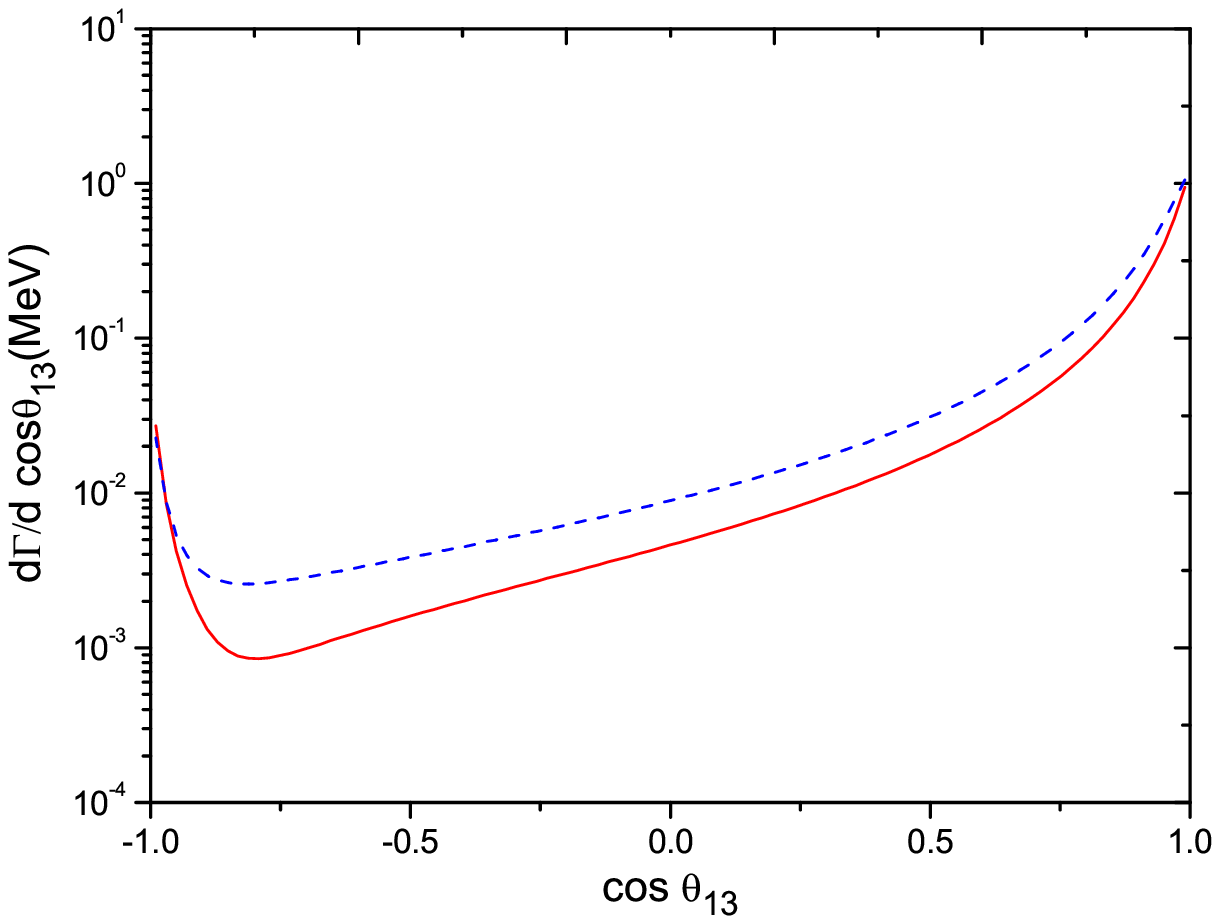}
\hspace{0.2cm}
\includegraphics[width=0.40\textwidth]{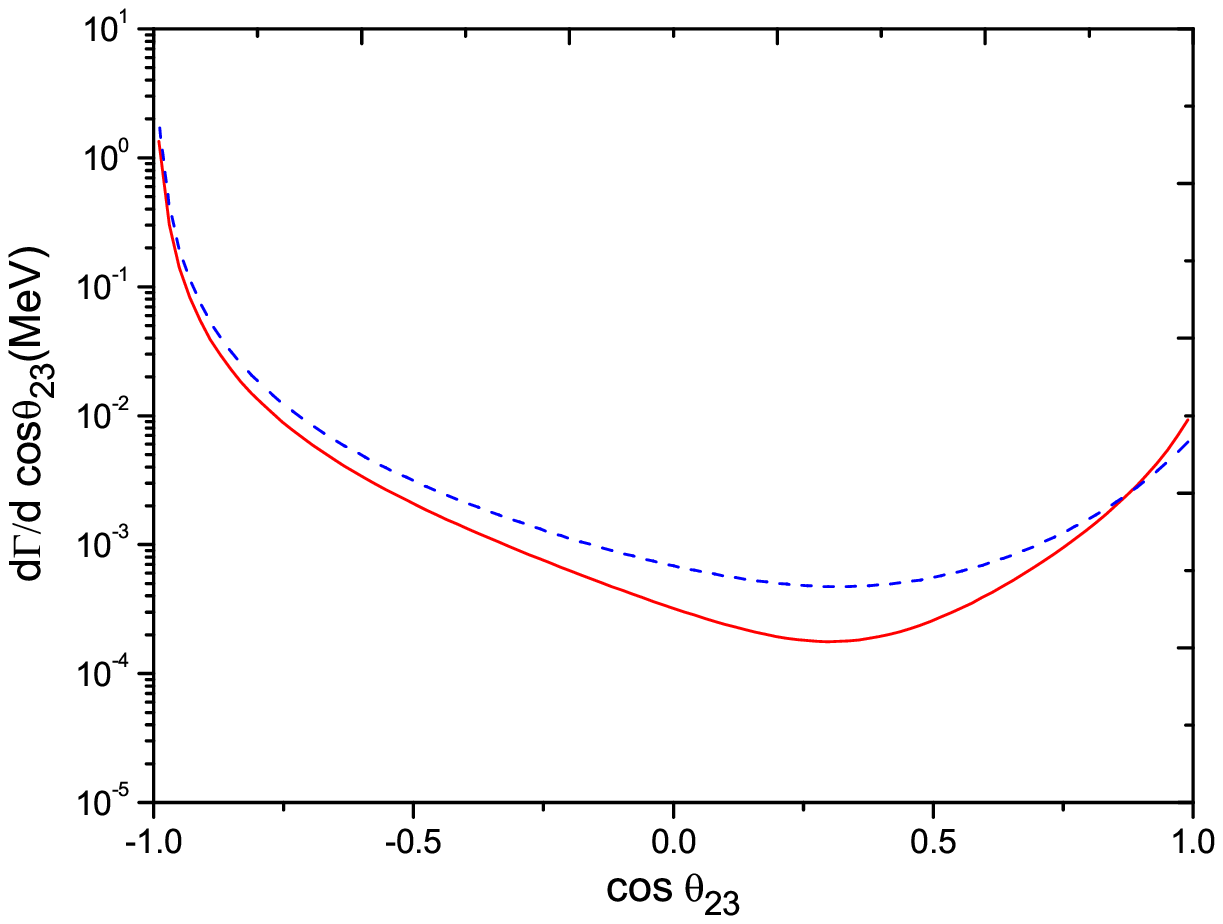}
\caption{Differential decay width $d\Gamma/d\cos{\theta_{13}}$ (Left) and $d\Gamma/d\cos{\theta_{23}}$ (Right) for the process $Z^0\rightarrow B^{(*)}_c +b+\bar c$, where the solid and the dashed lines are for $B_c$ and $B^*_c$, respectively. } \label{discos}
\end{figure*}

The differential distributions of the invariant masses $s_1$ and $s_2$, i.e. $d\Gamma/ds_1$ and $d\Gamma/ds_2$ are shown in Fig.(\ref{diss1s2}), where $s_1=(q_1+q_3)^2$ and $s_2=(q_1+q_2)^2$. And the differential distributions of $\cos\theta_{13}$ and $\cos\theta_{23}$, i.e. $d\Gamma/d\cos\theta_{13}$ and $d\Gamma/d\cos\theta_{23}$ are shown in Fig.(\ref{discos}), where $\theta_{13}$ is the angle between $\vec{q}_1$ and $\vec{q}_3$, and $\theta_{23}$ is the angle between $\vec{q}_2$ and $\vec{q}_3$ respectively. It can be found that the largest differential decay width of $d\Gamma/d\cos\theta_{13}$ is achieved when $\theta_{13}=0^{\circ}$, i.e. the $(c\bar{b})$-quarkonium and $c$-quark moving in the same direction. While the largest differential decay width of $d\Gamma/d\cos\theta_{23}$ is achieved when $\theta_{23}=180^{\circ}$, i.e. the $(c\bar{b})$-quarkonium and $b$-quark moving back to back.

Next, it would be interesting to show the theoretical uncertainties for the production. Main uncertainty sources include the matrix elements (or the wavefunction at the origin of the binding system $|\psi_{B^{(*)}_c}(0)|$), the renormalization scale $\mu_R$, the constitute quark masses $m_b$ and $m_c$. $|\psi_{B^{(*)}_c}(0)|$ and $\alpha(\mu_R)$ are overall parameters for the present case, and their uncertainties can be easily figured out. For example, one can set $\mu_R$ to be $2m_c$ or $2m_b$, since the intermediate gluon as shown in Fig.(\ref{feyn}) should be hard enough so as to produce a $c\bar{c}$-quark pair or a $b\bar{b}$-quark pair, which inversely ensures the pQCD applicability of the process. By setting these two scales to calculate the process, we obtain the ratio $\Gamma_{\mu_R=2m_b} / \Gamma_{\mu_R=2m_c}\propto\alpha^2_s(2m_b) / \alpha^2_s(2m_c)\sim 0.67$. In the following discussion, we fix $\mu_R=2m_c$ and $|\psi_{B^{(*)}_c}(0)|=0.361$ GeV$^{3/2}$ \cite{pot6}.

For clarity, we present the uncertainties of $m_c$ and $m_b$ in `a factorizable way'. When focussing on the uncertainties from $m_c$, we let it be a basic `input' parameter varying in a possible range $m_c=1.50\pm0.30\; {\rm GeV}$ with all the other factors, including the $b$-quark mass and {\it etc.} being fixed to their center values. Similarly, when discussing the uncertainty caused by $m_b$, we vary the $b$-quark mass $m_b$ within the region of $m_b=4.90\pm0.40\; {\rm GeV}$.

\begin{table}
\begin{tabular}{|c|c|c|c|}
\hline\hline $m_c$({\rm GeV})        & 1.20   & 1.50   & 1.80  \\
\hline $\Gamma_{(^1S_0)}({\rm KeV})$ & 183.5 & 81.4 & 42.2   \\
\hline $\Gamma_{(^3S_1)}({\rm KeV})$ & 280.1 & 116.4 & 57.1   \\
\hline
\end{tabular}
\caption{Decay width for the production of $B^{(*)}_c$ through $Z^0$ decay with varying $m_c$, where $m_b$ is fixed to be $4.9$ GeV. }
\label{tabmc}
\end{table}

\begin{table}
\begin{tabular}{|c|c|c|c|}
\hline\hline $m_b$ ( \ {\rm GeV})        & 4.50   & 4.90   & 5.30  \\
\hline $\Gamma_{(^1S_0)}({\rm KeV})$ & 82.1 & 81.4 & 71.0   \\
\hline $\Gamma_{(^3S_1)}({\rm KeV})$ & 114.1 & 116.4 & 95.6   \\
\hline
\end{tabular}
\caption{Decay width for the production of $B^{(*)}_c$ through $Z^0$ decay with varying $m_b$, where $m_c$ is fixed to be $1.5$ GeV. }
\label{tabmb}
\end{table}

The decay width for the production of $B^{(*)}_c$ through $Z^0$ decay with varying $m_c$ or $m_b$ are presented in TAB.\ref{tabmc} and TAB.\ref{tabmb}. It shows that the decay width is more sensitive to $m_c$, which decreases with the increment of $m_c$.

By adding these two uncertainties caused by $m_b$ and $m_c$ in quadrature, we obtain
\begin{equation}
\Gamma_{(^1S_0)}=81.4^{+102.1}_{-40.5} \;{\rm KeV}
\end{equation}
and
\begin{equation}
\Gamma_{(^3S_1)}=116.4^{+163.9}_{-62.8} \;{\rm KeV} .
\end{equation}

\begin{figure*}
\includegraphics[width=0.4\textwidth]{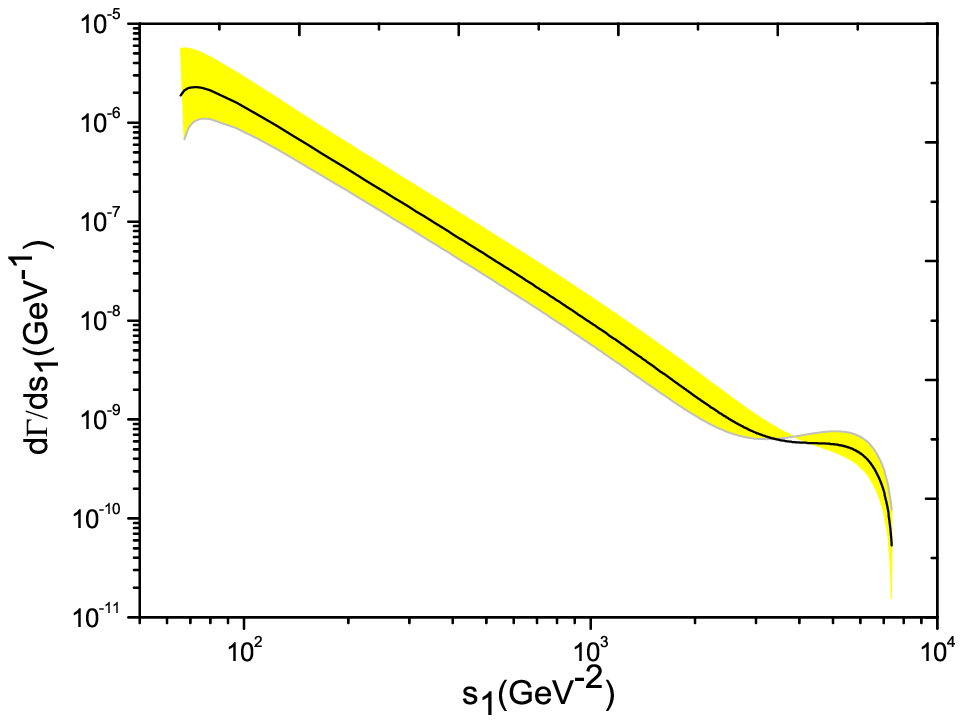}
\hspace{0.2cm}
\includegraphics[width=0.4\textwidth]{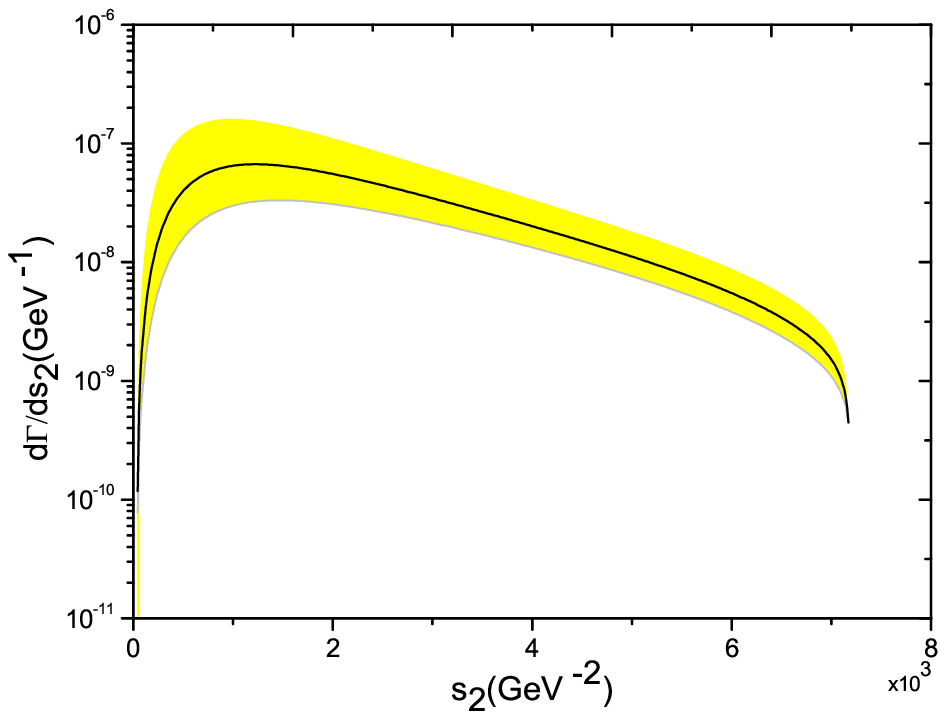}
\caption{Uncertainties of differential decay width $d\Gamma/ds_1$ (Left) and $d\Gamma/ds_2$ (Right) for $Z^0\rightarrow B_c +b+\bar c$, where the contributions from $^1S_0$ and $^3S_1$ are summed up. } \label{sun}
\end{figure*}

\begin{figure*}
\includegraphics[width=0.4\textwidth]{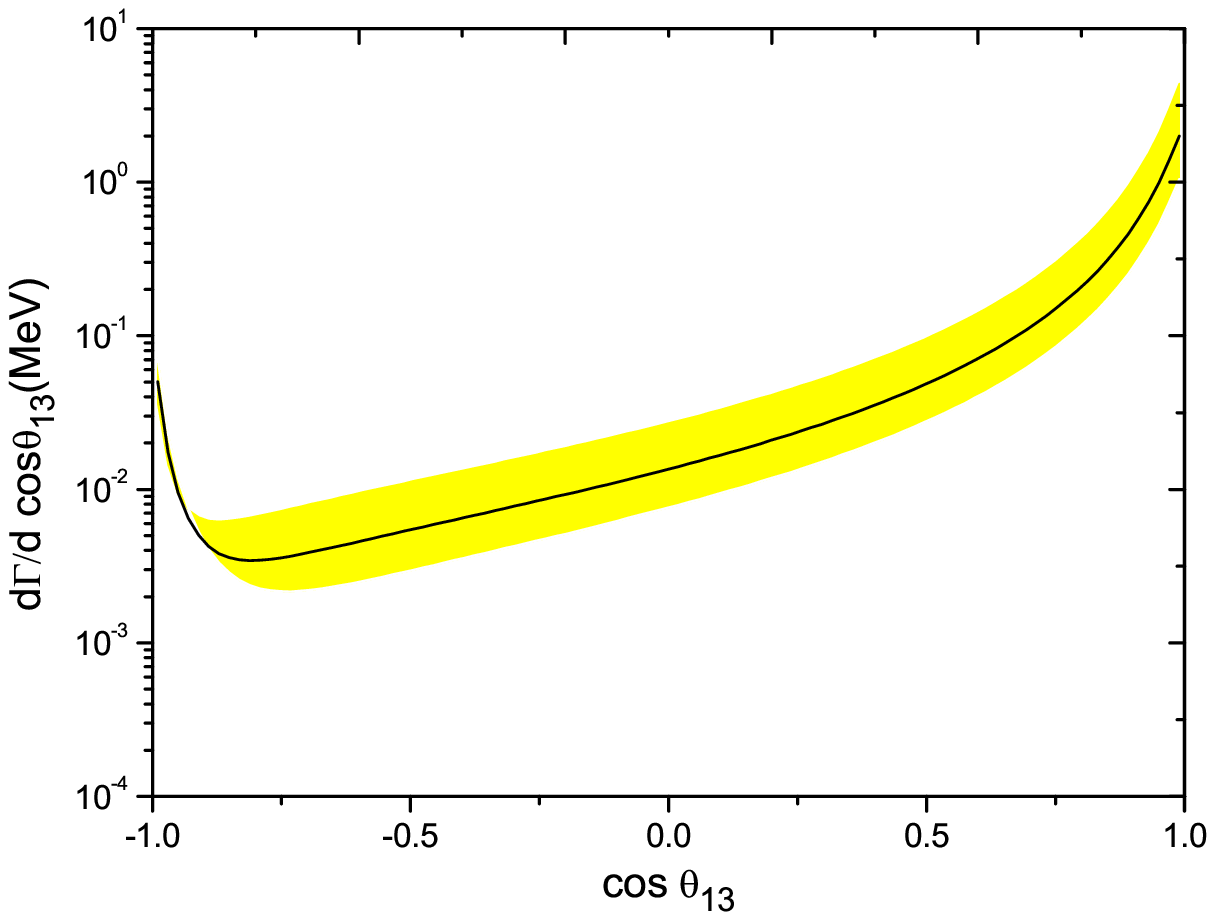}
\hspace{0.2cm}
\includegraphics[width=0.4\textwidth]{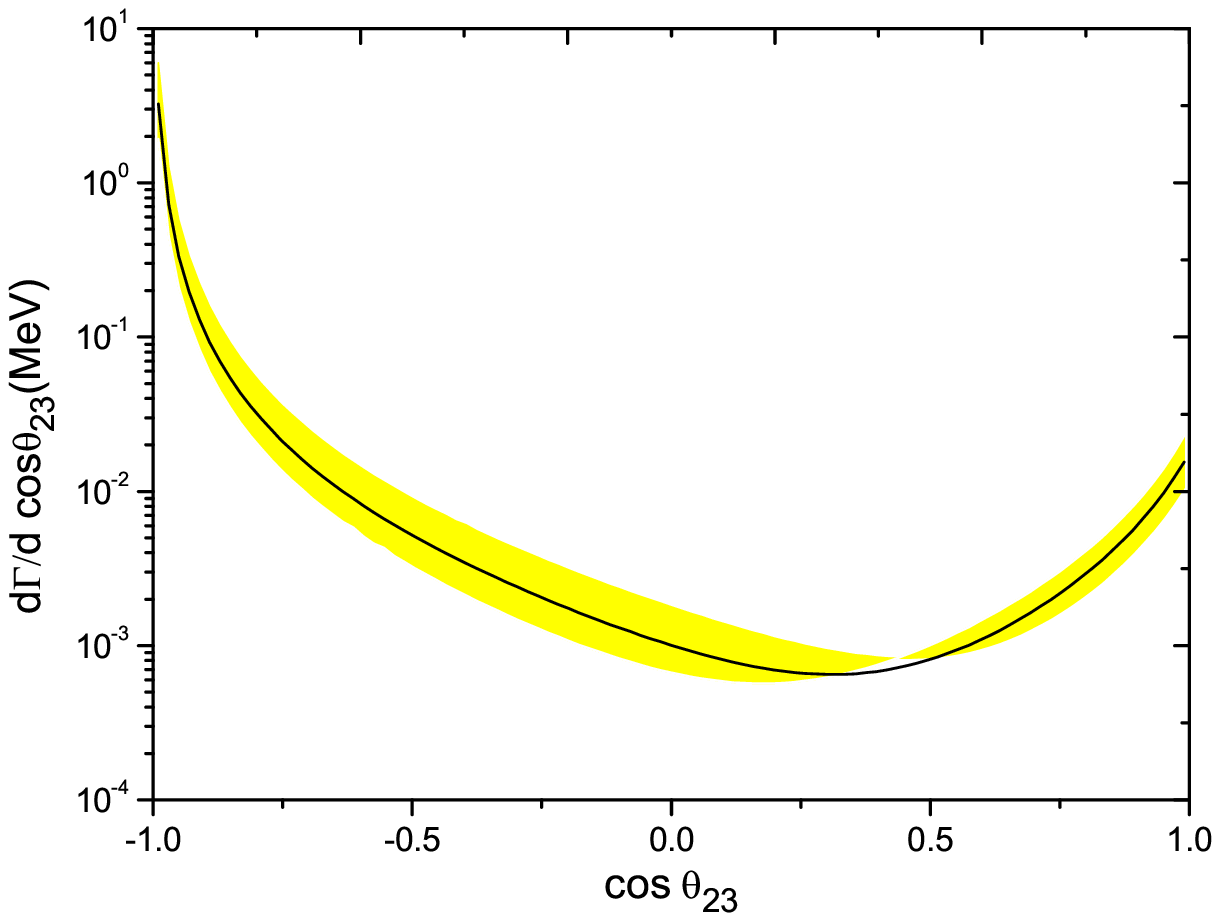}
\caption{Uncertainties of differential decay width $d\Gamma/d\cos{\theta_{13}}$ (Left) and $d\Gamma/d\cos{\theta_{23}}$ (Right) for $Z^0\rightarrow B_c +b+\bar c$, where the contributions from $^1S_0$ and $^3S_1$ are summed up. } \label{cosun}
\end{figure*}

The shaded bands in Figs.(\ref{sun},\ref{cosun}) show the corresponding uncertainty more clearly, where the contributions from $^1S_0$ and $^3S_1$ are summed up. The center solid line is for $m_c=1.5 {\rm GeV}$ and $m_b=4.9 {\rm GeV}$, the upper edge of the band is obtained by setting $m_c=1.2 {\rm GeV}$ and $m_b=5.3 {\rm GeV}$, while the lower edge of the band is obtained by setting $m_c=1.8 {\rm GeV}$ and $m_b=4.5 {\rm GeV}$.

As a summary: by using the `new trace amplitude approach', we calculate the $B_c$ production process, $Z^0\to B^{(*)}_c +b+\bar{c}$. The decay widths together with their uncertainties caused by the $b$ and $c$ quark masses are $\Gamma_{(^1S_0)}=81.4^{+102.1}_{-40.5}$ KeV and $\Gamma_{(^3S_1)}=116.4^{+163.9}_{-62.8}$ KeV, where the errors are caused by varying $m_b$ and $m_c$ within their reasonable regions $m_c\in[1.2, 1.8]$ GeV and $m_c\in[4.5,5.3]$ GeV. Further more, the differential decay width for $s_{1,2}$ and $\cos\theta_{13,23}$ together their uncertainties are drawn in Figs.(\ref{sun},\ref{cosun}). Considering the advantage of the clean environment in $e^{+}e^{-}$ collider, it will provide another useful platform in studying the $B_c$ production.

\hspace{2cm}

Acknowledgements: This work was supported in part by Natural Science Foundation Project of CQ CSTC under Grant No.2008BB0298, by Natural Science Foundation of China under Grant No.10805082 and No.11075225, and by the Fundamental Research Funds for the Central Universities under Grant No.CDJZR101000616.

\appendix

\section{Formulae for the phase space integration}

The decay width of the process $Z^0(k) \to B^{(*)}_c(q_3) + b(q_2) + \bar c(q_1)$ is proportional to the following phase space,
\begin{equation}
d\Gamma \propto \frac{(2\pi)^4}{2k^0}\prod\limits_{f = 1}^3 {d^3} {\vec{q}_f} \frac{\delta^4(k - \sum_{f=1}^3 {q_f})} {(2\pi)^3 2q_f^0}
\end{equation}
where $k=(k^0,\vec k)=(k^0,k^1,k^2,k^3)$, $q_f=(q_f^0,\vec q_f)=(q_f^0,q_f^1,q_f^2,q_f^3)$. Furthermore, in the rest frame of $Z^0$ boson ($k^0=m_Z$), we have
\begin{widetext}
\begin{eqnarray}
\frac{d\Gamma}{ds_1 ds_2}&\propto& \frac{1}{(2\pi)^5(2m_Z)} d^4q_1 d^4q_2d^4q_3 \delta(q_1^2-m_c^2) \delta(q_2^2-m_b^2) \delta(q_3^2-m_{B_c}^2)\theta(q_1^0)\theta(q_2^0)\theta(q_3^0)\nonumber\\
&&\times \delta(s_1-(q_1+q_3)^2)\delta(s_2-(q_1+q_2)^2)\delta(k-q_1-q_2-q_3)\nonumber\\
&\propto& \frac{1}{2^6 \pi^5 m_Z} d^4q_2d^4q_3\delta((k-q_2-q_3)^2-m_c^2) \delta(q_2^2-m_b^2)\delta(q_3^2-m_{B_c}^2) \theta(k^0-q_2^0-q_3^0)\nonumber\\
&&\times \theta(q_2^0)\theta(q_3^0)\delta(s_1-(k-q_2)^2)\delta(s_2-(k-q_3)^2)\nonumber\\
&\propto& \frac{1}{2^8 \pi^5 m^3_Z}d^3\vec{q}_2d^3\vec{q}_3 \delta(q_2^{0^2}-\vec{q}_2^2-m_c^2) \delta(q_3^{0^2}-\vec{q}_3^2-m_{B_c}^2) \theta(k^0-q_2^0-q_3^0)\nonumber\\
&&\times \theta (q_2^0)\theta (q_3^0)\delta(s_1+m_{B_c}^2-m_c^2-2m_Zq_3^0+2q_2^0q_3^0-2\vec{q}_2 \cdot \vec{q}_3)\nonumber\\
&\propto& \frac{|\vec{q}_2|\cdot|\vec{q}_3|}{2^{10} \pi^5 m_Z^3} d\Omega_2 \sin\theta_{23}d\theta_{23} d{\phi_{23}}\theta(k^0-q_2^0-q_3^0)\theta(q_2^0)\theta(q_3^0)\nonumber\\
&&\times \delta(s_1+s_2-m_Z^2-m_c^2+2q_2^0q_3^0-2|\vec{q}_2|\cdot|\vec{q}_3|\cos\theta _{23}) \nonumber\\
&\propto& \frac{1}{2^8 \pi^3 m^3_Z}\theta (k^0-q_2^0-q_3^0)\theta (q_2^0)\theta (q_3^0)\theta(X) ,
\end{eqnarray}
\end{widetext}
where $q_2^0=\frac{m_Z^2+m_b^2-s_1}{2m_Z}$ and $q_3^0=\frac{m_Z^2+m_{B_c}^2-s_2}{2m_Z}$, $|\vec q_2|= \sqrt{q_2^{0^2}-m_b^2}$ and $|\vec q_3|=\sqrt{q_3^{0^2}-m_{bc}^2}$. The step function $\theta(X)$ is determined by ensuring $|\cos\theta_{23}|\leq1$, where
\begin{displaymath}
\cos\theta_{23} =\frac{s_1 + s_2 - m_Z^2 - m_c^2 + 2q_2^0 q_3^0}{2\left|\vec{q}_2 \right| \left|\vec{q}_3\right|}.
\end{displaymath}
With all these step function above, we can get the integration ranges:
\begin{widetext}
\begin{eqnarray}
s_1^{\min } &=& m_c^2 + m_{B_c}^2 - \frac{\left(s_2 - m_Z^2 + m_{B_c}^2 \right) \left(s_2 - m_b^2 + m_c^2 \right) + \sqrt{\eta\left(s_2,m_Z^2,m_{B_c}^2\right)\eta\left(s_2,m_b^2,m_c^2 \right)}}{2s_2}\\
s_1^{\max } &=& m_c^2 + m_{B_c}^2 - \frac{\left( s_2 - m_Z^2 + m_{B_c}^2 \right) \left(s_2 - m_b^2 + m_c^2 \right) - \sqrt {\eta \left( s_2, m_Z^2, m_{B_c}^2\right) \eta\left(s_2,m_b^2,m_c^2 \right)}} {2s_2}\\
s_2^{\min} &=& \left(m_c + m_b \right)^2 \\
s_2^{\max} &=& \left(m_Z - m_{B_c} \right)^2
\end{eqnarray}
\end{widetext}
where $\eta(x,y,z)=(x-y-z)^2-4yz$.

Furthermore, we can obtain the $\cos\theta_{23}$ distribution
\begin{eqnarray}
\frac{d\Gamma}{ds_1 d\cos\theta_{23}} \propto \frac{J}{2^7 \pi^3 m_Z^3}\theta(k^0 - q_2^0 - q_3^0)
\theta (q_2^0)\theta (q_3^0)\theta (X)\nonumber\\
\end{eqnarray}
where the extra Jacobian
\begin{equation}
J=\frac{- \left| {{\vec q}_2} \right|\left| {\vec q}_3 \right|}{\left| {1 - \frac{q_2^0}{m_Z} + \frac{\left| {{{\vec q}_2}} \right|(m_Z^2 + m_{B_c}^2 - {s_2})}{m_Z \sqrt {m_{B_c}^4 - 2(m_Z^2 + {s_2})m_{bc}^2 + (m_Z^2 -s_2 )^2}}\cos\theta _{23}}\right|}
\end{equation}
and there are two $s_2$ in different range of $\cos\theta_{23}$ and $s_1$, i.e.
\begin{widetext}
\begin{eqnarray}
s^{\pm}_2=&&\frac{1}{|\vec{q_2}|^2\cos^2\theta_{23}-(q_2^0-m_Z)^2} \Bigg\{[|\vec{q_2}|^2(m_{B_c}^2 +{m_Z}^2)\cos^2\theta_{23}-(q_2^0-m_Z) [m_Z(s_1 +q_2^0 m_Z)+q_2^0 m_{B_c}^2-m_c^2-{m_Z}^2]] \nonumber\\
&&\pm m_Z|\vec{q_2}|\cos\theta_{23}[m_{B_c}^4-2m_{B_c}^2(m_c^2+2(m_Z-q_2^0)^2  -s_1-2|\vec{q_2}|^2 \cos^2\theta_{23})+(m_c^2-s_1)^2]^\frac{1}{2}\Bigg\} ,
\end{eqnarray}
\end{widetext}
where $s^{+}_2$ is obtained when $\cos\theta_{23}\in [0,-1]$ and $s_1\in [s_{1\min}[{\cos\theta_{23}}], s_{1\min}[{\cos\theta_{23}=0}]]$. And $s^{-}_2$ is obtained when $\cos\theta_{23}\in[1,0]$ and $s_1\in[s_{1\min}[\cos\theta_{23}=0],s_{1\max}]$ or $\cos\theta_{23}\in[0,-1]$ and $s_1\in[s_{1\min}[{\cos\theta_{23}}],s_{1\max}]$. The $\theta(X)$ function determines the boundary of $s_1$:
\begin{widetext}
\begin{eqnarray}
s_{1\max }&=&(m_Z  - m_b)^2 \\
s_{1\min}[\cos\theta_{23}]&=&\frac{m_b^{2} (\cos^2\theta_{23}-1) m_{B_c}^2 + m_Z^{2}(m_c^2 + m_{B_c}^2 \cos^2\theta_{23}) +m_{B_c} m_Z\sqrt{Y}} {(\cos^2\theta_{23}-1) m_{B_c}^2 +m_Z^2}
\end{eqnarray}
\end{widetext}
with
\begin{widetext}
\begin{eqnarray}
Y&=&4m_b^{2} m_{B_c}^2 \cos^4\theta_{23}- (m_b^4 +(6m_{B_c}^2 -2(m_c^2 +m_Z^2)) m_b^2 +((m_{B_c}-m_c)^2-m_Z^2) \nonumber\\
&&\times ((m_{B_c}+m_c)^2 -m_Z^2)) \cos^2\theta_{23} +\left(m_b^2 +m_{B_c}^2 -m_c^{2} -m_Z^2\right)^2.
\end{eqnarray}
\end{widetext}
The distribution for $\cos\theta_{13}$ can be obtained in a similar way.

\section{Amplitude of the process $Z^0(k)\rightarrow B^{(*)}_c(q_3) + b(q_2) +\bar c(q_1)$}

The amplitude $M$ of the process $Z^0(k)\rightarrow B^{(*)}_c(q_3) + b(q_2) +\bar c(q_1)$ has the general structure
\begin{equation}
M = {\bar u_s}({q_2})A{v_{s'}}({q_1}) ,
\end{equation}
where $A$ can be read from Eqs.(\ref{A1})-(\ref{A4}).

To derive analytical expression for the process and to make its form simpler as much as possible, we adopt the `new trace amplitude approach' suggested by Refs.\cite{chang1,tbc2} to do our calculation. Detailed process of the approach can be found in Refs.\cite{chang1,tbc2}, and here, we shall only list our main results.

After summing up the spin states, the square of the amplitude can be divided into four parts,
\begin{equation}
|M|^2 = |M_{1}|^2 + |M_{2}|^2 + |M_{3}|^2 + |M_{4}|^2,
\end{equation}
where by introducing a light-like momentum $k_0$ and a spacelike vector $k_1$ that satisfies the relations, $k_1\cdot k_1=-1$ and $k_0\cdot k_1=0$, the four amplitude $M_i$ can be written as
\begin{eqnarray}
M_1 &=& \frac{N}{{\sqrt 2 }} Tr\left[ {({\slashed{q}_1} - {m_c}){\slashed{k}_0}({\slashed{q}_2} + {m_b})A } \right] ,\nonumber\\
M_2 &=& \frac{N}{{\sqrt 2 }} Tr\left[ {({\slashed{q}_1} - {m_c}){\gamma _5}{\slashed{k}_0}({\slashed{q}_2} + {m_b})A } \right] , \nonumber\\
M_3 &=& \frac{N}{{\sqrt 2 }} Tr\left[ {({\slashed{q}_1} - {m_c}){\slashed{k}_0}{\slashed{k}_1}({\slashed{q}_2} + {m_b})A } \right]\nonumber
\end{eqnarray}
and
\begin{equation}
M_4 = \frac{N}{{\sqrt 2 }} Tr\left[ {({\slashed{q}_1} - {m_c}){\gamma _5}{\slashed{k}_1}{\slashed{k}_0}({\slashed{q}_2} + {m_b})A } \right] ,\nonumber
\end{equation}
where $N = 1/\sqrt {4({k_0}\cdot{q_1})({k_0}\cdot{q_2})}$ is the normalization constant.
$k_0$ and $k_1$ are arbitrary momenta, and in order to write down $M_n$ as explicitly and simply as possible: \\
1) We set $k_0 = {q_2} - \alpha {q_1}$, where the coefficient $\alpha$ is determined by the requirement that $k_0$ be a lightlike vector:
\begin{equation}
\alpha = \frac{{q_1} \cdot {q_2} \pm \sqrt{({q_1} \cdot {q_2})^2 - m_b^2m_c^2}}{m_c^2} .
\end{equation}
2) We set $k_1^\mu  = i{N_0}{\varepsilon ^{\mu \nu \rho \sigma }}{q_{1\nu }}{k_{\rho }}{q_{2\sigma }}$, where $N_0$ ensures $k_1\cdot k_1=-1$. It is found that $\slashed{k}_1$ can be expressed as,
\begin{equation}
\slashed{k}_1 = {N_0}{\gamma _5}\left[ {{q_1} \cdot {k}{\slashed{q}_2} + {\slashed{q}_1}{k} \cdot {q_2} - {q_1} \cdot {q_2}{\slashed{k}} - {\slashed{q}_1}{\slashed{k}}{\slashed{q}_2}} \right]. \\
\end{equation}
And then the resultant $M_i$ can be simplified as:	
\begin{eqnarray}
M_1 &=& {L_1} \times Tr [({\slashed{q}_1} - m_c)({\slashed{q}_2} + {m_b})A] ,\\
M_2 &=& {L_2} \times Tr [({\slashed{q}_1} - m_c){\gamma _5}(\slashed{q}_{2}+{m_b})A] ,\\
M_3 &=& M_{3'} - {N_0}[{m_b} ({q_1} \cdot {k}) + {m_c}({q_2} \cdot {k})]M_2 ,\\
M_4 &=& M_{4'} + {N_0}[{m_b} ({q_1} \cdot {k}) - {m_c}({q_2} \cdot {k})]M_1 ,
\end{eqnarray}
where
\begin{eqnarray}
M_{3'} &=&\frac{N_0}{4L_2} Tr\left[ {({\slashed{q}_1} - {m_c}){\gamma _5}{\slashed{k}}({\slashed{q}_2} + {m_b})A } \right] , \\
M_{4'} &=&-\frac{N_0}{4L_1} Tr\left[ {({\slashed{q}_1} - {m_c}){\slashed{k}}({\slashed{q}_2} + {m_b})A } \right] .
\end{eqnarray}

Furthermore, the amplitudes $M_i$ can be expanded over some basic Lorentz structures:
\begin{equation}
M_i(n)=\sum^m_{j=1} A^i_j(n) B_j(n) (i=1-4)
\end{equation}
and
\begin{equation}
M_{i'}(n)=\sum^m_{j=1} A^{i'}_j(n) B_j(n) \;\; (i'=3,4)
\label{amat}
\end{equation}
where $m$ is the number of basic Lorentz structure $B_j(n)$, whose value dependents on the $(c\bar{b})$-quarkonium state $n$: e.g.
$m=3$ for $n=(c\bar{b})[^1S_0]_1$, $m=12$ for $n=(c\bar{b})[^3S_1]_1$. As for $A^3_j(n)$ and $A^4_j(n)$, they can be expressed by
\begin{eqnarray}
A^3_j(n) &=& A^{3'}_j(n)-{N_0}[{m_b} ({q_1} \cdot {k}) + {m_c}({q_2} \cdot {k})] A^2_j(n) , \nonumber\\
A^4_j(n) &=& A^{4'}_j(n)+{N_0}[{m_b} ({q_1} \cdot {k}) - {m_c}({q_2} \cdot {k})] A^1_j(n) . \nonumber
\end{eqnarray}
The explicit expression for $A^{1,2}_j(n)$ and $A^{3',4'}_j(n)$ of each state shall be listed in the following subsections.

To shorten the notation,  we set $T_b=\frac{1}{4}-\frac{1}{3}{\sin ^2}{\theta _w}$ and $T_c=\frac{1}{4}-\frac{2}{3}{\sin ^2}{\theta _w}$. And, we define some dimensionless parameters
\begin{displaymath}
r_1=\frac{m_b}{m_Z},\;\; r_2=\frac{m_c}{m_Z},\;\;
r_3=\frac{m_{B_c}}{m_Z}
\end{displaymath}
and
\begin{eqnarray}
&&x=q_3\cdot k/m_Z^2=\frac{1}{2m_Z^2}(m_{B_c}^2+m_Z^2-s_2),\nonumber\\
&&y=q_2\cdot k/m_Z^2=\frac{1}{2m_Z^2} (m_b+m_Z^2-s_1), \nonumber\\
&&z=q_1\cdot k/m_Z^2=\frac{1}{2m_Z^2}(m_Z^2+m_c^2-s_3), \nonumber\\
&&u=q_3\cdot q_2/m_Z^2 =\frac{1}{2m_Z^2}(s_3-m_{B_c}^2-m_b^2), \nonumber\\
&&v=q_3\cdot q_1/m_Z^2=\frac{1}{2m_Z^2}(s_1-m_{B_c}^2-m_c^2), \nonumber\\
&&w=q_1\cdot q_2/m_Z^2= \frac{1}{2m_Z^2}(s_2-m_b^2-m_c^2),\nonumber
\end{eqnarray}
where $s_1=(q_1+q_3)^2$, $s_2=(q_1+q_2)^2$, and $s_3=(q_2+q_3)^2$, which satisfy the relation: $s_1+s_2+s_3=m_Z^2+m_c^2+m_b^2+m_{B_c}^2$. And the short notations
for the denominators are
\begin{eqnarray}
&&d_1=\frac{1}{(q_2-k)^2-m_b^2}\frac{1}{(q_{31}+q_1)^2},\nonumber\\
&&d_2=\frac{1}{(k-q_{32})^2-m_b^2}\frac{1}{(q_{31}+q_1)^2},\nonumber\\
&&d_3=\frac{1}{(q_{32}+q_2)^2}\frac{1}{(q_{31}-k)^2-m_c^2},\nonumber\\
&&d_4=\frac{1}{(q_{32}+q_2)^2}\frac{1}{(q_3+q_2)^2-m_c^2},\nonumber
\end{eqnarray}
Furthermore, the following relations are useful to short the expressions:
\begin{displaymath}
u+v+r_3^2=x,\;\; w+u+r_2^2=y, \;\; w+v+r_1^2=z.
\end{displaymath}

\subsection{Coefficients for the production of $B_c$}

There are 3 basic Lorentz structures $B_j$ for the case of $B_c$ ($^1S_0$), which are
\begin{displaymath}
B_1=\frac{q_3\cdot\epsilon(k)}{m_Z} ,\; B_2=
\frac{q_2\cdot\epsilon(k)}{m_Z},\; B_3=\frac{i}{m_Z^3}\varepsilon(k,q_3,q_2,\epsilon(k)),
\end{displaymath}
where $\varepsilon(k,q_3,q_2,\epsilon(k))=\varepsilon^{\mu\nu\rho\sigma}
k_{\mu}q_{3\nu} q_{2\rho} \epsilon_\sigma(k)$. The values of the
coefficients $A^{1}_j$ and $A^{3'}_j$ are
\begin{widetext}
\begin{eqnarray}
A^1_1 &=&\frac{L_1 m_Z{}^{7/2} }{\sqrt{r_3}}\Bigg((r_1 (1-2 r_1 r_3)-(2 r_1- r_3) y)d_1+(r_1(r_3+2 r_3 u-2 r_1 x)-r_3^2 y)d_2 +(r_1 (2 r_2 (x-u)+r_3 (y-1))\nonumber\\
&&+r_2 (r_3 y-2 r_2 u))d_3+(r_1^3+2 (r_2-2 r_3) r_1^2 +(r_2^2-r_3^2-y) r_1+(r_2-2 r_3) (2 u-y))d_4\Bigg),\\
A^1_2 &=&\frac{L_1 m_Z{}^{7/2} }{\sqrt{r_3}}((r_3 (x-2 u)+r_3 (4 u-2 x-4 y+2))d_1+(r_3 x-2 r_1r_3^2)d_2   \nonumber\\
&&+ (2 r_2 r_3^2-r_3 x)d_3+(-2 r_1 r_3^2+2 r_2 r_3^2-2 u r_3-x r_3)d_4),\\
A^1_3 &=&-4 L_1 m_Z{}^{7/2} \sqrt{r_3} T_b(d_1+d_2+d_3+d_4),\\
A^{3'}_1 &=&\frac{m_Z{}^{9/2} N_0} {{L_2} r_3^{3/2}}(T_b (-2 y^2+y-r_1 r_3)r_3 d_1
+T_b (-2 x r_1^3 +(r_3 (4 x+4 y-3)-2 r_2 x) r_1^2 +(r_2 r_3-2 u+2 (r_3^2- \nonumber\\
&& 2 r_2r_3 +2 u) y) r_1+r_3 (2 u+(-2 x-2 y+1) y))d_2 -T_c ((r_3-2 r_2 x) r_1^2+r_2 (r_3 (4 x+4 y-3) -2 r_2 x) r_1 \nonumber\\
&& -2 r_3 y^2-2 r_2 u+(-4 r_3 r_2^2+2 r_3^2 r_2+4 u r_2+r_3) y)d_3+T_c (r_1 r_3 (1-2 x) +(2r_3(r_3 -3r_1)-2 u) y)r_3 d_4),\\
A^{3'}_2 &=&\frac{m_Z{}^{9/2} N_0} {L_2 \sqrt{r_3}}(T_b (2 x r_1^2+2 (-2 x r_3-2 y r_3+r_3+r_2x)r_1 +(-r_3^2+2 r_2 r_3-2 u+x) \times(2 y-1))d_1 \nonumber\\
&&-T_b (r_3^2+x (-2 x-2 y+1))d_2-T_c (r_3^2+x (2 y-1))d_3 +T_c (x r_1^2+2 (-2 x r_3-2 y r_3+r_3+r_2 x) r_1 \nonumber\\
&&+r_2^2 x+2 r_2 r_3 (2y-1) -(r_3^2+2 u) (x+2 y-1))d_4),\\
A^{3'}_3 &=&\frac{m_Z{}^{9/2} N_0 }{4 L_2 \sqrt{r_3}}((2 r_1 r_3+2 y-1)d_1+(2 r_1 r_3-2 x-2 y+1)d_2-(2r_2 r_3+2 y-1)d_3 + (2 r_3^2-4 r_2 r_3+2 u)d_4)
\end{eqnarray}
The values of the coefficients $A^{2}_j$ and $A^{4'}_j$ are
\begin{eqnarray}
A^2_1 &=&-\frac{4 L_2 m_Z{}^{7/2} }{\sqrt{r_3}}(T_b ((2r_1+ r_3) y-r_1)d_1+T_b ((5 r_1-r_2)
r_3 y-r_1 (2 r_1 r_3^2+(4 r_1 (r_1-r_2)+4 u \nonumber\\
&&+1) r_3+2r_1 u-2 r_2 u-2 r_1 x))d_2+(r_1 r_2 r_3 T_c(6r_1-2r_2)+r_3 T_c r_1+(4 r_2 r_1+2 r_2 r_3 )T_c u    \nonumber\\
&&-2 r_2 T_c x r_1-r_3 T_c y r_1-3 r_2 r_3 T_c y)d_3+(-r_3 T_c r_1^3+4 r_3^2 T_c r_1^2+r_3^3 T_c r_1 -4 r_2 r_3^2 T_c r_1     \nonumber\\
&&+r_2^2 r_3 T_c r_1+r_3 T_c y r_1+4 r_3^2 T_c u-2 r_2 r_3 T_c u-2 r_3^2 T_c y+r_2 r_3 T_c y)d_4),\\
A^2_2 &=&-\frac{4 L_2 m_Z{}^{7/2} }{\sqrt{r_3}}(T_b ((6r_1-2r_2)r_1 r_3
+(4 u-2 x-4 y+2) r_3 +2 r_1 u-2 r_2 u-3 r_1 x+r_2 x)d_1  \nonumber\\
&&+(r_2-r_1) T_b x d_2+(r_1-r_2) T_c x d_3-T_c (-2 r_3 ((r_1-r_2)^2-r_2 r_3) -2 (r_1-r_2) u-r_3 x)d_4),\\
A^2_3 &=&\frac{L_2 m_Z{}^{7/2} }{\sqrt{r_3}}( r_3 d_1+(r_2-r_1)d_2+(r_2-r_1)d_3-r_3 d_4),\\
A^{4'}_1 &=&\frac{m_Z{}^{9/2} N_0}{4 L_1 r_3{}^{3/2}}((r_1 r_3+y) (2 y-1)d_1
+(2 x r_1^3-(r_3+2 r_2 x) r_1^2+(r_2 r_3+2 u-4 u y) r_1 \nonumber\\
&&-r_3(2 u+(-2 x-2 y+1) y))d_2+((r_3-2 r_2 x) r_1^2+r_2 (2 r_2 x-r_3) r_1    \nonumber\\
&&+(2 y-1) (2 r_2 u-r_3 y))d_3+(r_1 r_3 (2 x-1)-(-4 r_1 r_3-2 u) y)d_4),\\
A^{4'}_2 &=&\frac{m_Z{}^{9/2} N_0}{4 L_1 \sqrt{r_3}}( (r_3 (2 r_1-2 r_2+r_3)+2 u-x) (2 y-1)d_1+(r_3^2+2 r_1 (x+2 y-1) r_3+x \nonumber\\
&&-2 x (x+y))d_2+(r_3^2-2 r_2 (x+2 y-1) r_3+x (2 y-1))d_3+(-x r_1^2+2 r_3 (2 x+2 y-1) r_1 \nonumber\\
&&+r_2^2 x+(r_3^2+2 u) (x+2 y-1)-2 r_2 r_3 (2 x+2y-1))d_4),\\
A^{4'}_3 &=&-\frac{ N_0 m_Z{}^{9/2}}{L_1 \sqrt{r_3}} (T_b (2 y-1) d_1+T_b (-2 x-2 y+1) d_2 +(T_c-2T_c y)d_3+T_c (r_1^2-r_2^2+r_3^2+2 u)d_4)
\end{eqnarray}
\end{widetext}

\subsection{Coefficients for $B^*_c$}

There are 12 basic Lorentz structures $B_j$ for the case of
$B^*_c$ $(^3S_1)$, which are
\begin{eqnarray}
B_1 &=& \epsilon(k)\cdot\epsilon(q_3),\;\;
B_2 =\frac{i}{m_Z^2}\varepsilon(k,q_3,\epsilon(k),\epsilon(q_3)),\nonumber\\
B_3 &=&\frac{i}{m_Z^2}\varepsilon(k,q_2,\epsilon(k),\epsilon(q_3)),\;\;
B_4 = \frac{i}{m_Z^2}\varepsilon(q_3,q_2,\epsilon(k),\epsilon(q_3)),\nonumber\\
B_5 &=&\frac{k\cdot\epsilon(q_3) q_3\cdot\epsilon(k)}{m_Z^2},\;\;
B_6 =\frac{k\cdot\epsilon(q_3) q_2\cdot\epsilon(k)}{m_Z^2},\nonumber\\
B_7 &=&\frac{q_2\cdot\epsilon(q_3) q_3\cdot\epsilon(k)}{m_Z^2},\;\;
B_8 =\frac{q_2\cdot\epsilon(k) q_2\cdot\epsilon(q_3)}{m_Z^2}, \nonumber\\
B_9 &=&\frac{i}{m_Z^4} \varepsilon(k,q_3,q_2,\epsilon(k))k \cdot \epsilon(q_3),\nonumber\\
B_{10} &=& \frac{i}{m_Z^4} \varepsilon(k,q_3,q_2,\epsilon(q_3))q_3 \cdot \epsilon(k), \nonumber\\
B_{11} &=& \frac{i}{m_Z^4} \varepsilon(k,q_3,q_2,\epsilon(q_3))q_2 \cdot \epsilon(k), \nonumber\\
B_{12} &=& \frac{i}{m_Z^4} \varepsilon(k,q_3,q_2,\epsilon(k))q_2 \cdot \epsilon(q_3) . \nonumber
\end{eqnarray}
The values of the coefficients $A^{1}_j$ and $A^{3'}_j$ are
\begin{widetext}
\begin{eqnarray}
A^1_1 &=&\frac{4 L_1 m_Z{}^{7/2} }{\sqrt{r_3}}(T_b (r_2 y+r_1 (x+y-1))r_3 d_1+T_b (-2 xr_1^2+(r_3 (x-1)-2 r_2 x) r_1  \nonumber\\
&&-2 u x+r_3^2 y+2 x y)d_2-T_c (2 x r_1^2+(2 r_2 x+r_3 (x-1)) r_1+2 u x+r_3^2 y-2 x y)d_3  \nonumber\\
&&-T_c (r_1^3-(r_2^2-r_3^2-2u+x+y) r_1+r_2 y)r_3 d_4),\\
A^1_2 &=&\frac{L_1 m_Z{}^{7/2} }{\sqrt{r_3}}(-r_1 r_3 d_1+(r_1 (-2r_1- r_3)-4 u+4y)d_2 +(2 r_1^2+r_3 r_1+4 u-4 y)d_3-r_1 r_3 d_4),\\
A^1_3 &=&\frac{L_1 m_Z{}^{7/2} }{\sqrt{r_3}}((r_1-r_2)r_3 d_1+(3 r_3^2-2 x)d_2-(r_3^2-2x)d_3- (r_1-r_2) r_3 d_4),\\
A^1_4 &=&-\frac{2 L_1 m_Z{}^{7/2} }{\sqrt{r_3}}(r_1 r_3 d_1+(r_1 r_3+x-1)d_2+(r_2
r_3-x+1)d_3+r_2 r_3 d_4),\\
A^1_5 &=&\frac{4 L_1 m_Z{}^{7/2} }{\sqrt{r_3}}(-r_1 r_3 T_b d_1+T_b (2 r_1^2+r_3 r_1+2 u-2 y)d_2 +T_c(2 r_1^2+r_3 r_1+2 u-2 y)d_3- r_1 r_3 T_c d_4),\\
A^1_6 &=&-4 L_1 m_Z{}^{7/2}( (3 r_1+r_2) \sqrt{r_3} T_b d_1+r_3^{3/2} T_b d_2-r_3^{3/2}
T_c d_3+(r_1-r_2) \sqrt{r_3} T_c d_4),\\
A^1_7 &=&8 L_1 m_Z{}^{7/2}(-r_1 \sqrt{r_3} T_b d_2+r_2 \sqrt{r_3} T_c d_3+r_3^{3/2}T_c d_4),\\
A^1_8 &=&8 L_1 m_Z{}^{7/2} r_3^{3/2}(T_b d_1+T_c d_4),\\
A^1_9 &=&A^1_{10}=\frac{2 L_1 m_Z{}^{7/2}}{\sqrt{r_3}}(d_2-d_3),\\
A^1_{11} &=&A^1_{12}=0,\\
A^{3'}_1 &=&-\frac{m_Z{}^{9/2} N_0 }{4 L_2 \sqrt{r_3}}(r_3 ((2 x+2 y-1) r_1^2+(r_2-2 r_2 y) r_1+u+y-2 y (x+y))d_1 +(-2 x r_1^3+(r_3 (4 x+4 y-3)\nonumber\\
&&-2 r_2 x) r_1^2+(r_2 r_3-2 (u+(x-1) x)+2 (r_3^2-2 r_2r_3+2 u-x) y) r_1+2 r_2 x y+r_3 (-2 y^2-2 x y+y+u))d_2   \nonumber\\
&&-((r_3-2 r_2 x) r_1^2+(-2 x r_2^2+r_3 (4 x+4 y-3) r_2-2 x (x+y-1)) r_1 -2 r_2u-4 r_2^2 r_3 y+ 2 r_2 (r_3^2+2 u+x) y\nonumber\\
&&+r_3 (-2 y^2-2 x y+y+u))d_3  +r_3 (-u-r_1 (r_1+r_2 (2 x-1))+((r_1-r_2)^2+r_3^2+2 u) y)d_4),\\
A^{3'}_2 &=&\frac{2 m_Z{}^{9/2} N_0 }{L_2 \sqrt{r_3}}(T_b (r_1 (x-1)+(r_1 -r_2) y)d_2 -r_3 T_b y d_1 +T_c (r_1 (x-1)+(r_1 -r_2) y)d_3-r_3 T_c (r_1^2+u)d_4),\\
A^{3'}_3 &=&\frac{m_Z{}^{9/2} N_0 \sqrt{r_3} }{L_2}(T_b (1-2 y)d_1+T_b (2 x+2 y-1)d_2+T_c(2 x+2 y-1)d_3 +2r_2 r_3 T_c d_4),\\
A^{3'}_4 &=&-\frac{m_Z{}^{9/2} N_0 \sqrt{r_3} }{L_2}(T_b (d_1+d_2)+T_c d_3+T_c (2 x+2
y-1) d_4),\\
A^{3'}_5 &=&\frac{m_Z{}^{9/2} N_0}{2 L_2 \sqrt{r_3}}(r_3 (r_1^2-y)d_1+(r_2 y-r_1(u+x+y-1))d_2 + (r_3 y-r_2 (u+y)+r_1 (x+y-1))d_3+r_3(r_1^2+u)d_4),\\
A^{3'}_6 &=&\frac{m_Z{}^{9/2} N_0 \sqrt{r_3}}{4 L_2}( - (2 u+2 y-1)d_1+(2 r_1 r_3-2 x-2 y+1)d_2 +(2r_2 r_3+2 y-1)d_3+2r_2 r_3 d_4),\\
A^{3'}_7 &=&-\frac{m_Z{}^{9/2} N_0 }{4 L_2 \sqrt{r_3}}(-r_3 d_1+(r_3-2 r_1 x)d_2+(r_3-2 r_2 x)d_3-r_3 (2 x+4 y-1)d_4),\\
A^{3'}_8 &=&\frac{m_Z{}^{9/2} N_0 \sqrt{r_3} }{2 L_2}(x+2 y-1)(d_1+d_4),\\
A^{3'}_9 &=&-\frac{2 m_Z{}^{9/2} N_0 \sqrt{r_3} }{L_2}T_c d_3,\\
A^{3'}_{10} &=&A^{3'}_{11}=\frac{2 m_Z{}^{9/2} N_0 }{L_2 \sqrt{r_3}}(r_1 T_b d_2- r_2 T_c d_3),\\
A^{3'}_{12} &=&-\frac{2 m_Z{}^{9/2} N_0 \sqrt{r_3} }{L_2}T_c d_4,
\end{eqnarray}
The values of the coefficients $A^{2}_j$ and $A^{4'}_j$ are
\begin{eqnarray}
A^2_1 &=&-\frac{L_2 m_Z{}^{7/2} }{\sqrt{r_3}}(r_3 (2 r_1^3-2 r_2 r_1^2+(2 u-x-3 y+1) r_1+r_2 y)d_1+(r_3 r_1(6r_1^2-2r_1 r_2)  \nonumber\\
&&+2 (u-2 x) r_1^2+(2 r_2 (x-u)+r_3 (4 u-x-5 y+1)) r_1-2u x+r_2 r_3 y+2 x y)d_2  \nonumber\\
&&-((4 r_2 r_3-2 x) r_1^2+(2 r_2 r_3^2-(4 r_2^2+x+y-1) r_3+2 r_2 u)r_1-2 r_2^2 u+r_2 r_3 (4 u-3 y)  \nonumber\\
&&+2 x (y-u))d_3+r_3(r_1^3-2 r_2 r_1^2+(r_2^2+r_3^2+2 u-x-y) r_1+r_2 (y-2 u))d_4),\\
A^2_2 &=&\frac{4 L_2 m_Z{}^{7/2} }{\sqrt{r_3}}(-r_1 r_3 T_b d_1+T_b (r_1(4 r_1-r_3)+4 (u-y))d_2 +T_c (r_1 (4 r_1-r_3)+4 (u-y))d_3-r_1 r_3 T_c d_4),\\
A^2_3 &=&\frac{4 L_2 m_Z{}^{7/2} }{\sqrt{r_3}}(- (r_1-r_2)r_3 T_b d_1+T_b (2x-2r_2-r_3)d_2 +T_c (2x-2r_2-r_3)d_3-(r_1-r_2) r_3 T_c d_4),\\
A^2_4 &=&\frac{8 L_2 m_Z{}^{7/2} }{\sqrt{r_3}}(T_b (x-1)d_2+T_c (x-1)d_3),\\
A^2_5 &=&-\frac{L_2 m_Z{}^{7/2}}{\sqrt{r_3}}( r_1 r_3 d_1+(r_1 (4 r_1- r_3)+2(u-y))(d_2-d_3)+ r_1 r_3 d_4),\\
A^2_6 &=&-L_2 m_Z{}^{7/2} (r_1-r_2) \sqrt{r_3}(d_1+d_2-d_3+d_4),\\
A^2_7 &=&2 L_2 m_Z{}^{7/2} r_1 \sqrt{r_3}(d_1+d_4),\\
A^2_8 &=&2 L_2 m_Z{}^{7/2} (r_1-r_2) \sqrt{r_3}(d_1+d_4),\\
A^2_9 &=&A^2_{10}=-\frac{8 L_2 m_Z{}^{7/2} }{\sqrt{r_3}}(T_b d_2+T_c d_3),\\
A^2_{11} &=&A^2_{12}=0,\\
A^{4'}_1 &=&\frac{m_Z^{9/2} N_0}{L_1 \sqrt{r_3}}(r_3 T_b (r_1^2-r_2 r_1+u+(-2 x-2 y+1) y)d_1 +T_b (r_3 r_1^2+(2 x (x+y-1)-r_2 r_3) r_1 +2 r_2 x y \nonumber\\
&&+r_3 (y+u-2 y^2-2 x y))d_2 +T_c ((r_2 r_3+2 x (x+y-1)) r_1-r_3 r_1^2+2 r_2 x y-r_3 (y+u-2 y^2-2 x y))d_3    \nonumber\\
&&+r_3 T_c ((y-1) r_1^2+r_2 r_1-u+(-r_2^2+r_3^2+2 u) y)d_4),\\
A^{4'}_2 &=&\frac{m_Z{}^{9/2} N_0 }{2 L_1 \sqrt{r_3}}( r_3 (y-r_1^2)d_1+(r_1 (r_1^2-r_2r_1+2 u+x-1)+(2 r_2-r_3) y)d_2  \nonumber\\
&&+ (r_2 r_1^2-(r_2^2+x-1) r_1+2 r_2 u-(2 r_2+r_3) y)d_3+r_3 (r_1^2-r_2 r_1+u)d_4),\\
A^{4'}_3 &=&\frac{m_Z{}^{9/2} N_0}{4 L_1 \sqrt{r_3}}(r_3 (2 r_1 (r_2-r_1)+2 y-1)d_1+ (-2 r_1r_3^2-2 y r_3+r_3+2 (r_1-r_2) x)d_2  \nonumber\\
&&+ (-2 r_2 r_3^2-2 y r_3+r_3+2 (r_2-r_1)x)d_3-r_3 (r_3^2-(r_1-r_2)^2)d_4),\\
A^{4'}_4 &=&\frac{m_Z{}^{9/2} N_0 }{4 L_1 \sqrt{r_3}}(r_3 d_1+(2 r_2-r_3+4 r_1 (x+y-1))d_2   +(2 r_1-r_3+4 r_2 (x+y-1))d_3+r_3 (2 x+2 y-1)d_4),\\
A^{4'}_5 &=&-\frac{2 m_Z{}^{9/2} N_0}{L_1 \sqrt{r_3}}( -r_3 T_b y d_1+T_b (r_1 (r_1^2-r_2r_1+u+x-1)+r_3 y)d_2+T_c (-r_2 r_1^2  \nonumber\\
&&+(r_2^2+x+y-1) r_1-r_2 u+(r_2+r_3)y)d_3-r_3 T_c (r_1^2-r_2 r_1+u)d_4),\\
A^{4'}_6 &=&\frac{m_Z{}^{9/2} N_0 \sqrt{r_3} }{L_1}(T_b (2 r_1 (r_1-r_2)+2 u+2 y-1)d_1+T_b(2 x+2 y-1)d_2 +(T_c-2 T_c y)d_3-4r_1 r_2 T_c d_4),\\
A^{4'}_7 &=&\frac{m_Z{}^{9/2} N_0}{L_1 \sqrt{r_3}}( r_3 (-T_b d_1 +T_b d_2+T_c d_3) +(2 x+4 y-1)(2T_b r_1 d_2 -2T_c r_2 d_3 -T_c r_3 d_4)),\\
A^{4'}_8 &=&-\frac{2 m_Z{}^{9/2} N_0 \sqrt{r_3} }{L_1}(x+2 y-1)(T_b d_1+T_c d_4),\\
A^{4'}_9 &=&-\frac{m_Z{}^{9/2} N_0 }{2 L_1 \sqrt{r_3}}(2r_1 d_2-(r_1-r_2)d_3),\\
A^{4'}_{10} &=&-\frac{m_Z{}^{9/2} N_0 }{2 L_1 \sqrt{r_3}}(r_1 d_2+r_2 d_3),\\
A^{4'}_{11} &=&\frac{m_Z{}^{9/2} N_0 \sqrt{r_3}}{2 L_1}d_1,\\
A^{4'}_{12} &=&\frac{m_Z{}^{9/2} N_0 \sqrt{r_3}}{2 L_1}d_4 .
\end{eqnarray}
\end{widetext}

\end{document}